\documentclass[aps,pra,twocolumn,superscriptaddress]{revtex4-2}
\usepackage{physics}
\usepackage{color}
\usepackage{amsmath}
\usepackage{algorithm}
\usepackage{algpseudocode}
\usepackage[normalem]{ulem}
\usepackage{bm}
\usepackage{physics}
\usepackage{amsfonts}
\usepackage{mathrsfs}
\usepackage{lipsum}
\usepackage{graphicx}

\begin{document}


\title{Adiabatic quantum learning}


\author{Nannan Ma}
\affiliation{Department of Physics, National University of Singapore, Singapore 117551, Singapore}
\author{Wenhao Chu}
\affiliation{Department of Physics, National University of Singapore, Singapore 117551, Singapore}

\author{Jiangbin Gong}
\email{phygj@nus.edu.sg}
\affiliation{Department of Physics, National University of Singapore, Singapore 117551, Singapore}
\affiliation{Centre for Quantum Technologies, National University of Singapore, Singapore 117543, Singapore}
\affiliation{Joint School of National University of Singapore and Tianjin University, International Campus of Tianjin University, Binhai New City, Fuzhou 350207, China}


\date{\today}

\begin{abstract}
Adiabatic quantum control protocols have been of wide interest to quantum computation due to their robustness and insensitivity to their actual duration of execution.   As an extension of previous quantum learning algorithms, this work proposes to execute some quantum learning protocols based entirely on adiabatic quantum evolution, hence dubbed as ``adiabatic quantum learning".    In a conventional quantum machine learning protocol, the output is usually the expectation value of a pre-selected observable and the projective measurement of which forces a quantum circuit to run many times to obtain the output with a reasonable precision.  By contrast, the proposed adiabatic quantum learning here may be integrated with future adiabatic weak measurement protocols, where a single measurement of the system allows to extract the expectation value of observables of interest without disrupting the concerned quantum states.  Our main idea is illustrated with simple examples.  

\end{abstract}


\maketitle

\section{Introduction}

Quantum computation \cite{lloyd1996universal, divincenzo1995quantum} is expected to lead the next computation revolution by fully exploiting quantum advantages \cite{nielsen2002quantum}.   The converged efforts on scalable quantum computation would not be possible by pioneering theoretical work \cite{shor1999polynomial, grover1996fast, montanaro2016quantum,liu2021rigorous} and experimental demonstrations \cite{peruzzo2014variational, kandala2017hardware, vandersypen2001experimental,arute2019quantum}. Meanwhile, inspired by the capability of machine learning \cite{carleo2019machine}, there have been tremendous efforts to transplant machine learning thinking and  algorithms to  quantum platforms \cite{biamonte2017quantum, cong2019quantum, schuld2019quantum}, hence the promising research field of quantum machine learning (QML). To date QML is mainly based on the so-called variational quantum circuits (VQCs)\cite{mcclean2016theory,yuan2019theory, goto2021universal,mitarai2018quantum,havlivcek2019supervised, lloyd2020quantum, schuld2020circuit}. In VQCs, the output, namely, the expectation value of an observable of interest, is normally obtained from many conventional, hence strong and projective measurements.  For example, an initial state (e.g., the vacuum state $ \ket{0} $) goes through a series of quantum parameterized gate operations (unitary operators).  These parameterized gates are introduced to encode $N$-dimensional input $ \textbf{x}=(x_1,...,x_N) $ as well as $M$ training variables $ \textbf{w}=(w_1,...,w_M) $. A prediction $ f_{\textbf{w}}(\textbf{x}) $ can be made from  the estimation $ \bra{0}U^\dagger(\textbf{x},\textbf{w})\hat{O}U(\textbf{x},\textbf{w})\ket{0}$ of a projective measurement on observable $ \hat{O} $, where $U^\dagger(\textbf{x},\textbf{w})$ is the overall time evolution operator.  As a stimulating development, a quantum version of the so-called universal approximation theorem (UAT)\cite{goto2021universal, perez2021one} has been obtained, confirming the expressivity of a quantum circuit to approximate any integrable functional relationships.  

What motivated this work is the recognition that the above outlined architecture of VQCs still relies on the necessary repetition of many projective measurements \cite{von2018mathematical} in extracting the expectation value of an observable of interest. The precision of measurement is inherently related with the variance of the eigenvalues of a concerned observable.  In situations involving excitations in a high-dimensional Hilbert space, the convergence of the measured observable can be slow.  Even for a low-dimensional Hilbert space,  a large number of quantum gates are still involved to encode many parameters, and hence it is highly costly to repeat the entire sequence of quantum gate operations after a projective measurement.   Moreover, if non-commuting observables are both of interest,  extra measurements must be performed by restarting the learning algorithm all over again.  That a measurement collapses and hence destroys the quantum state under measurement is not an issue of VQCs themselves. Yet, to find ways to mitigate this issue would constitute an important progress \cite{alonso2022single,banchi2021measuring,huang2020predicting,gresch2023guaranteed,kohda2022quantum}.   

The real question of this work is hence the following: Is it possible to execute QML algorithms in a way that is compatible with some innovative measurement protocols, so that we can avoid the repetition of costly quantum gate operations \cite{solinas2023quantum}? 
Our proposal of adiabatic quantum learning here aims to offer a positive answer, at least theoretically.
Indeed, we advocate that in future quantum learning algorithms, the innovate adiabatic weak measurement protocol \cite{aharonov1988result, AHARONOV199338} may replace projective measurements so as to save tremendously the needed number of quantum gate operations in implementing QML. In a nutshell, adiabatic quantum measurement is based on the weak coupling between a quantum system and a measuring device, say,  $ H_{\mathrm{int}}=\frac{\hat{p}\hat{O}}{T} $, over a sufficiently long integration time $T$, where $\hat{p}$ is the momentum operator of the pointer.  A simple consideration of the first-order perturbation theory indicates that if the system is prepared on one of its energy eigenstates, then there will be a small energy shift 
$\frac{p \langle \hat{O}\rangle} {T}$.  Assuming that quantum adiabatic conditions \cite{born1928beweis,kato1950adiabatic}  are obeyed, the system will stay on its instantaneous eigenstates. Meanwhile, the energy shift of the system-pointer coupling causes the pointer to change over a distance proportional to $ \expval{O} $. Throughout this process, the system and the pointer will not get entangled due to the perturbative nature of the system-pointer coupling.  Indeed the adiabatic condition ensures that the system remains at its slightly perturbed energy eigenstates.  This being the case, upon finally checking the actual position shift of the pointer and subject to the initial spreading of the pointer states as the only source of error without a lower bound \cite{zhang2020dissipative}, one can infer the expectation value of $\expval{O} $ without collapsing the measured state to one of the eigenstates of $\hat{O}$.   As is now clear, the readout of the said expectation value is achieved without incurring a wavefunction collapse.  As strange as it may sound,  recent brilliant experiments have demonstrated the in-principle feasibility of such a weak measurement protocol \cite{piacentini2017determining,pan2020weak}. It is important to note that
the prerequisite to implement this weak measurement scheme is to prepare the system under measurement on one of its energy eigenstates. As a matter of fact, it is the Hamiltonian of the system itself 
that protects the system on its initial eigenstate when a weak coupling between the system and a pointer is turned on. 

To show that quantum learning algorithms can be compatible with the above-summarized weak adiabatic measurement scheme,  in the following we shall depict a learning architecture which is based entirely on adiabatic quantum gates, that is, throughout the process the system is always on one of its energy eigenstates.  This way, in principle weak quantum adiabatic measurement can be introduced at any step of  a quantum learning algorithm without destroying the quantum state, and the expectation value of an observable of interest can be extracted in principle, without the need to repeat the quantum gate sequences. The extraction of non-coummuting observables, as pointed by Anharonov's paper itself \cite{aharonov1988result}, is also possible and hence potentially brings in even more resource saving.  This paper is organized as follows. we will illustrate first (i) how to encode a learning algorithm into operations on instantaneous Hamiltonians, (ii) how to design operations to make adiabatic quantum evolution possible (iii) how to execute the training procedure for such architectures.  We shall present two working examples where our protocol is applied to conventional classification tasks.  

\section{Theoretical considerations}

\subsection{Representation and operations}

Consider a quantum system described by a $D$-dimensional Hilbert space. In general, the complete set of all possible observables include $ D^2 $ elements, one of them being the identity and the rest can be all traceless and Hermitian. Because the identity element does not affect the eigenvalues (gap) and the eigenstates of the possible system Hamiltonian, we may omit this operator and hence only keep $ D^2-1 $ elements. These element will be denoted as $ e_i, i\in{1,...,D^2-1} $, forming the lie algebra of lie group SU(D) , with known and important properties such as $ [e_a, e_b]=2i\sum_{c=1}^{D^2-1}C_{abc}e_c $,  where $ C_{abc} $ is real and called the structure constant.  The structure constant is not unique, but we can always find the proper choice of $ e_i $, so that $ \mathrm{Tr}(e_ae_b)=2\delta_{ab} $ and $ C_{abc}=\frac{1}{4i}\mathrm{Tr}([e_a, e_b]e_c) $.  It is obvious that the structure constant is anti-symmetric in all three indices and must be 0 once two indices are the same.

A time-dependent Hamiltonian $ H(t) $, which will be necessary in our quantum learning algorithms, will be characterized with a $ d=D^2-1 $ dimensional unit vector $ \vec{n}(t) $, namely, the Hamiltonian vector. Under the assumption that initial system stays at the ground state of $ H(\vec{n}(0)) $ and the ensuing evolution is adiabatic, the final state of the system must be the ground state of $ H(\vec{n}(T)) $. What the final Hamiltonian is depends on the evolution path of the instantaneous Hamiltonians, or more precisely the track of $ \vec{n}$ in the $d$-dimensional parameter space of the Hamiltonian.  As such, it is intuitive to incorporate the learning process into the track of $ \vec{n}$ in terms of time.   As explained in Introduction, we start from an eigenstate of the Hamiltonian and attempt to ensure adiabatic time evolution afterwards is to make sure that at any point of the quantum manipulation, the state is the eigenstate of the instantaneous Hamiltonian and as such, it is possible to use the above-mentioned weak adiabatic measurement scheme to probe the system without destroying the current quantum state.  

To ensure that the time evolution of the system can be adiabatic, we need to make sure that along the track of $ \vec{n}$ in time, the eigenvalues of the time-evolving Hamiltonians do not touch.   This will restrict all possible time dependence of the Hamiltonian to a small subset.  We adopt the following simple strategy.   Let a Hamiltonian be mapped to a point of the $d$-dimensional unit sphere, only the rotation operation around a certain axis in the parameter space is considered.  That is,  in order to keep the energy-level gaps open during the entire quantum control protocol, only a subgroup of $ \mathrm{SO}(D^2-1) $ will be selected as our operators. One subgroup satisfying our conditions consists of all the rotations around the $d$ independent  axes. More technical details are presented in Appendix A, where it is explicitly shown that such kind of changes to the Hamiltonian can always be described by a unitary transformation of the Hamiltonian matrix, which can preserve the spectrum of the initial Hamiltonian and hence will not introduce any level crossing during the operation. Furthermore, as shown in Appendix A, a unitary transformation to a Hamiltonian can be always be understood as a rotational operation defined here applied to a Hamiltonian matrix. This indicates that the expressivity of our protocol based on a subset of rotational operations is as powerful as the conventional quantum circuit.

Much similar to our conventional understanding of the rotational group $\mathrm{SO}(3)$, one can identify a $ d $-dimensional unit vector $ \vec{m} $ as the rotation axis, and then set the rotation angle to be $ \theta $, so as to construct the operation ``rotation around $ \vec{m} $ by $ \theta $ ". The resulting rotation matrix can be expressed $ e^{\theta A} $, with $A$ being the generator corresponding to a rotation axis $ \vec{m} $ and is given by
\begin{equation}
	A_{ij}=\sum_{k}C_{ijk}m_k.
\end{equation}
Here  A is skew symmetric with $ A^T=-A $.

In a 2-dimensional Hilbert space (as an example, also to be used later), basic elements of a Hamiltonian are the three Pauli matrices $ \sigma_i $, $ i\in{1,2,3} $. They form the set of bases to depict an arbitrary  $ 2^2-1=3 $ dimensional unit vector $ \vec{n} $ to characterize a Hamiltonian: $ H(\vec{n})= \vec{n}\cdot\vec{\sigma} $. The structure constant is $ \epsilon_{abc} $ (Levi-Civita symbol) with $ [\sigma_a, \sigma_b]=2i\epsilon_{abc}\sigma_c $. Its rotation generator around $ \vec{m} $ is $ A_{ij}=\sum_{k}\epsilon_{ijk}m_k $ and corresponding rotation matrix is given by $ e^{\theta A}=\cos\theta I+\sin\theta A +(1-\cos\theta)A^2 $. The explicit expression of this rotational matrix is as the following:
\begin{widetext}
	\begin{equation}
		\label{rm}
		\scriptsize{
			\left[\begin{array}{ccc}
				\cos(\theta)+(1-\cos(\theta))m_1^2 & -\sin(\theta)m_3+(1-\cos(\theta))m_1m_2 & \sin(\theta)m_2+(1-\cos(\theta))m_1m_3 \\
				\sin(\theta)m_3+(1-\cos(\theta))m_2m_1 & \cos(\theta)+(1-\cos(\theta))m_2^2 & -\sin(\theta)m_1+(1-\cos(\theta))m_2m_3 \\
				-\sin(\theta)m_2+(1-\cos(\theta))m_3m_1 & sin(\theta)m_1+(1-\cos(\theta))m_3m_2 & 
				\cos(\theta)+(1-\cos(\theta))m_3^2
			\end{array}\right]}
	\end{equation}
\end{widetext}
Upon this rotation, the new unit vector depicting the new Hamiltonian becomes
\begin{equation}
		\vec{n'}=\cos(\theta)\vec{n}-\sin(\theta)\vec{n}\times\vec{m}+(1-\cos(\theta))(\vec{m}\cdot\vec{n})\vec{m}
\end{equation}
where $ \vec{n} $ and $ \vec{n'} $ are the Hamiltonian vector respectively before and after a rotation operator. For a 2-dimensional Hilbert space, these operations are the SO(3) rotational group. 
\begin{figure*}[htbp]
	\includegraphics[scale=0.25]{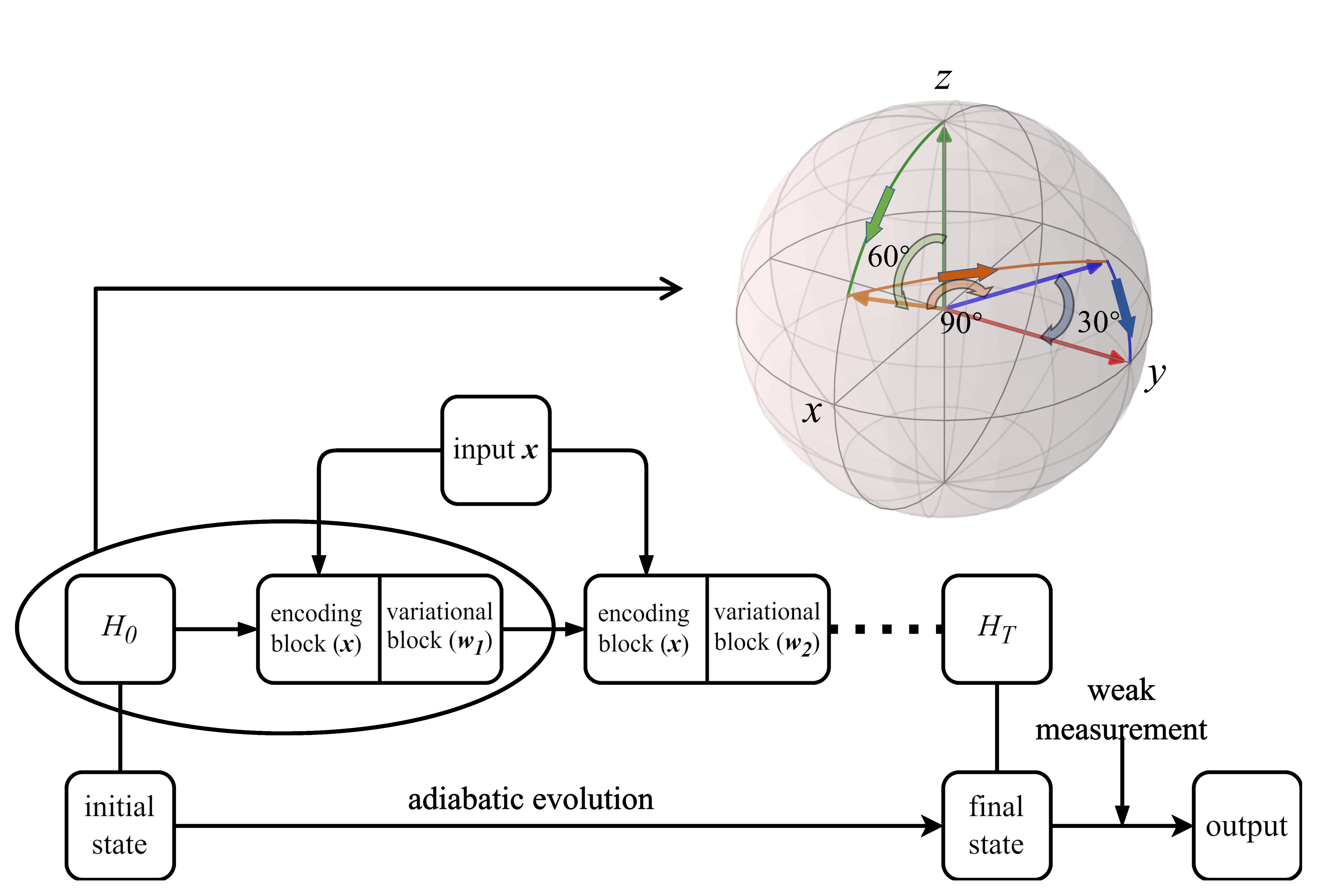}
	\caption{\label{f1}{\bf The flow of adiabatic learning protocol.} The initial Hamiltonian $ H_0 $ and initial state are prefixed. Input $x$ and variational parameters $w$ control the operations on the Hamiltonian as a function of time, and the evolution path of the Hamiltonian yields the adiabatic state evolution.   During the entire learning process, the system's state remains on one of its energy eigenstates. This allows our learning protocol to be integrated with weak adiabatic measurement, such that the expectation value of an  observable can be obtained without  collapsing the state of the system. The repetition of a learning unit improves the expressivity, just as how data re-uploading improves the learning capability of a conventional quantum circuit structure. For a 2-dimensional Hilbert space with 1-dimensional input $x=\pi/3$ and weight $ w_1=[\pi/2,\pi/6] $, there are three adiabatic  rotations respectively around the $y$ axis (in encoding clock) by $x$, around the $z$ axis (in variational block) by $  w_{11}$ and around the $x$ axis (in variational block) by $ w_{12} $. The Bloch sphere illustrates the track traced out by the Hamiltonian  starting from $ H_0=\sigma_z $. }
\end{figure*}
\subsection{Learning structure}

Our choice of a subset of operations to the system Hamiltonian is to make sure that adiabatic time evolution can be possible. Plus if this operation is indeed executed slowly,  then  adiabatic time evolution can be achieved throughout.  Under this assumption, the track of instantaneous Hamiltonian determines the output of a learning process. In our protocol, the ground state of the initial Hamiltonian $ H_0 $ is used as the initial state. They are hence fixed and easy to prepare. The classical input data is then encoded as the rotation angle with certain pre-selected axis of the parameter space.  This hence forms the encoding part.  Meanwhile, some operations are reserved as the adjustable part, usually the actual rotation angles along a certain axis of the parameter space, thus serving as the variational parameters. 

In principle, we can arrange rather arbitrarily the execution of the encoding step and the variational parameters in a learning structure. However, it is wise to exploit some useful concepts widely used in QVC in our adiabatic protocol. In particular,
let us choose a special operation configuration as an encoding block with their rotation parameters $ x $ as the input data, then another configuration as the variational block with their rotation parameters $ w $ as the variational parameters for the training step. These two blocks make up a unit.  {Given the mapping between the adiabatic evolution under consideration and possible unitary operations from a quantum circuit, many techniques that may improve the learning ability or expressivity can be transplanted into the execution of our learning protocol. One important technique is the so-called re-uploading strategy \cite{perez2020data}.  This strategy is to overcome the fact that a learned map $ f_\textbf{w}(\textbf{x})$ typically has a restricted frequency spectrum in its Fourier expansion.  According to several theoretical studies \cite{perez2021one, schuld2021effect}, the repetition of data uploading or adding more encoding layers can however expand its frequency spectrum and hence enable our learning model to gain a better learning capability.  Inspired by this progress, we propose to repeat our learning unit constructed above  several times, so as to also implement the data re-uploading strategy. For different units, their encoding blocks can be chosen to be the same, but parameters of different variational blocks are independent. For an input, we get a track of the unit vector representing the instantaneous Hamiltonian through a series of rotation operations determined by input and variational parameters. To drive an adiabatic evolution following this track, we introduce a  time scale, namely, $ g= \Delta_{t}/\Delta_{\theta} $,  where $\Delta_t$ is the duration to rotate an angle $\Delta_{\theta}$.  $g$ is hence the period needed for a unit change of rotation angle occurs.  The inverse of $g$  determines the speed of the adiabatic time evolution, which should be set to be sufficiently small.    The layout of this learning protocol is shown in Fig.~\ref{f1}. {To visualize the track of time-evolving Hamiltonian, the Bloch sphere in Fig.~\ref{f1} illustrates the effect of the first learning unit. The encoding block only has one rotation operation around the $y$-axis and the variational block consists of two operations respectively around the $z$ and $x$-axis, all preserving the spectrum of the Hamiltonian.  The initial Hamiltonian is $ \sigma_z $. For one-dimensional input $ x=\pi/3 $ and two-dimensional weight $ w_1=[\pi/2,\pi/6] $, the initial green vector is rotated around $y$ by $ 60^\circ $ to the brown intermediate vector, then around $z$ by $ 90^\circ $ to  the blue vector, finally around $x$ by $ 30^\circ $ to the red vector. As a result, the track is comprised of three arcs (green, brown and blue) depicting the adiabatic evolution path.}  At the end of such adiabatic evolution, the state is still the ground state of final Hamiltonian $ H_T $ where $ T $ is the total operation time.  This makes it possible to extract the expectation value of some observation by use of the weak adiabatic measurement advocated above. 

\subsection{Training}

Similar to other learning structures, an appropriate loss function is needed and it reflects the difference between predictions and targets. One must attempt to minimize this loss function in order to get optimal learning parameters in these variational blocks. 
In our training step, we assume that we can achieve perfectly adiabatic evolution. That is,  the time evolving state is always the ground state of the instantaneous Hamiltonian along its track.  As shown in the Appendix A, a rotation operation on the Hamiltonian is equivalent to such a unitary operation acting on its ground state. This indicates that there is a mapping between our adiabatic protocol and a conventional VQC, with our protocol adopting a special class of rotation operations that preserves the spectrum of the Hamiltonian.  To us this is good news since all currently known optimization methods applicable to a quantum circuit can be implemented in our adiabatic protocol as well.  There are three main methods based on gradient descent: automatic differentiation, numerical differentiation and parameter shift \cite{li2017hybrid,mitarai2018quantum}. Automatic differentiation (AD) is the most efficient one if the process is simulated on a classical computer. But if we do not wish to depend on a classical computer, then parameter shift is favored because it enables us to obtain the theoretical gradient through two measurement results. Numerical differentiation is the common choice to get an approximate result by perturbing the respective parameter. Note that if a method needs to have the expectation value from experiments to optimize, then a large number of measurements will be executed in traditional protocols.  This is no longer the case here since by construction, the expectation value of concerned observables can be obtained, at least in the future, by invoking weak adiabatic measurement protocols.   There are still many gradient-free optimization methods, which are however not considered in this work.  

		
		


Summarizing this subsection on the training step,  we make use of the availability of various libraries for building and training a quantum circuit in order to demonstrate that our idea can work in principle. To that end we  map our rotation operations to a quantum circuit and use available libraries to determine the optimized unitary transformations.  These unitary transformations can be regarded as rotations on the Hamiltonians.   If we now execute the rotations adiabatically,  the system always stays at the ground state of of the instantaneous Hamiltonian. The time evolving state of the system is hence subject to
the desired unitary operations determined from the circuit optimization (apart from a global phase).   
With this understanding, we can apply all optimization methods before we do the actual numerical simulation based on finite-time rotations of the Hamiltonian parameters.  There will be some errors introduced due to the absence of perfect adiabaticity. 
 However, our working examples below show that this can be a small issue.

\section{Results}

To demonstrate the feasibility of our adiabatic learning protocol, we examine its implementation in two binary classification tasks, one treats a one-dimensional data set and the other treats a two-dimensional data set. In the execution of training on a quantum circuit in the cases below, automatic differentiation is selected as the optimization method to get the optimal parameters. The actual AD optimizer is COBYLA.

\subsection{Case I}

As a benchmark step, we start with a working example about one-dimensional binary classification. The distribution of label is shown in Fig.~\ref{f2}(a). Input is a real number $ x $. There are three isometric parts in the range $ x\in(-1,1) $. The outer two parts (blue) are arranged with the same label $ y=1 $ and the central part (green) goes with the label $ y=0 $. 

To execute quantum learning, we next execute adiabatic learning protocol based on a two-dimensional Hilbert space, equivalent to a one-qubit system. The initial state under consideration is the qubit ground state $ \ket{0} $ and the initial Hamiltonian is $ -\sigma_z $, whose vector representation $ \vec{n}_0 $ is $(0,0,-1)$ according to our notation used in the previous section.  The unit vector is allowed to rotate on the Bloch sphere embedded in a three-dimensional parameter space.  In a learning unit, the encoding block only contains one rotation operation because of this one-dimensional input.  This operation can be chosen as the rotation around the $x$ axis whose unit vector is $(1,0,0)$. The rotation angle is based on the input $ x $. The variational block consists of three rotation operations, rotation around the $z$ axis $(0,0,1)$ by $ w_1 $, rotation around the $y$ axis $(0,1,0)$ by $ w_2 $ and rotation around the $z$ axis $(0,0,1)$ by $ w_3 $.  The variational parameter in this block is hence a 3-dimensional vector ($ w_1, w_2, w_3 $). To achieve a good precision, we repeat this learning unit for three times. For a given input, the track of the instantaneous Hamiltonian $ H(t) $ in time is determined by a total of $4\times 3=12$ rotations. This track of the qubit Hamitonian guides the evolution of the initial state, with the final state given by $ \ket{\phi_T}=\hat{T}\left\{\exp\left[-i\int_{0}^{T}dtH(t)\right]\right\}\ket{0} $, where $\hat{T}$ is the time ordering operator.  The adjustable parameters to train in the training step are implemented by three three-dimensional vectors, one each from one learning unit. From the adiabatic theorem, the final state is expected to be an eigenstate of the final instantaneous Hamiltonian $ H(T)$. The final outcome can be the expectation value of the observable $ \sigma_z $, denoted by $ e $, with $ e $  ranging from -1 to 1. The classification criterion is:
\begin{equation}
	y=
	\begin{cases}
		0 \quad e\leq0\\
		1 \quad e>0
	\end{cases}
\end{equation}
However, note that our implementation equally applies if the expectation value of any other osbervable $\sigma_x$ or $\sigma_y$ is used.   As mentioned before, in our actual numerical experiments, we transform these rotation operations to its quantum circuit analogy.  We then use the package Qiskit to construct the corresponding quantum circuit and train it with a selected optimizer to get these optimal parameters.

\begin{figure}[htbp]
	\includegraphics[scale=0.55]{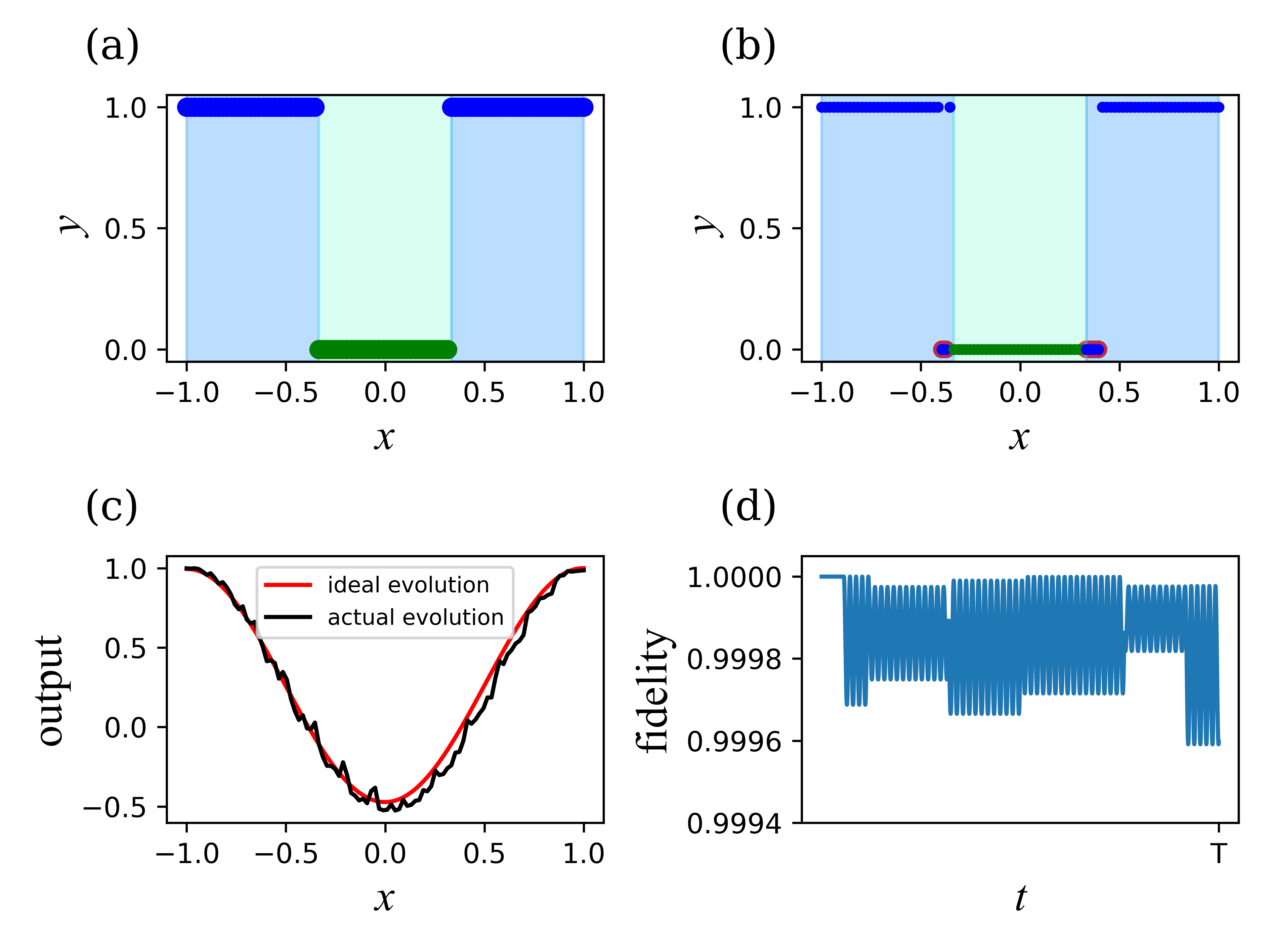}
	\caption{\label{f2} One-dimensional binary classification. (a) The label $ y $ distribution in range $x\in[-1, 1]$. $ y=1$ when $|x|>\frac{1}{3} $ and $y=0 $ otherwise. (b) The classification result of 100 samples. The misclassifications marked with red circles occur in the neighborhood of the two boundaries. The accuracy of the classification is $ 94\% $. (c) The output difference between ideal evolution with perfect adiabaticity and the actual time evolution in our numerical simulation.  The roughness of the actual simulation result is due to nonadiabatic effects. (d) The fidelity of evolution state against the eigenstate of instantaneous Hamiltonian for sample $ x=0 $ during the whole adiabatic evolution. Note that though the fidelity keeps at almost unity throughout the operation, the influence of finite-time operation on adiabaticity is still visible in panel (c). }
\end{figure}

According to distribution presented in Fig.~\ref{f2}(a), the train data consists with $ N_t = 20$ randomly selected samples. After training, we get optimal parameters, a 9-dimensional vector [-0.572, 0.643, 0.478, 1.57, 1.886, -1.225, -1.4, -1.568, 0.856]. The training score that represents the ratio of getting the correct prediction is 0.95, which means that there are only 1 sample classified wrongly. Then we use the test data set that contains 100 samples uniformly in range $ [-1, 1] $ to verify our adiabatic learning structure. Again, for a given input, the initial state $ \ket{0} $ evolves following the track of the instantaneous Hamiltonians.  We then get the output. In this process, we set the operation time-scale parameter $g=0.01/0.0005=20$ per angle change (the large this number is, the closer to the adiabatic limit). Then, the evolution period for different input will { be determined with $ \Delta_{t}=g\times \Delta_{\theta}$ (as mentioned in learning structure part)}.
 The final classification result is presented in Fig.~\ref{f2}(b), where the accuracy is seen to be at  94\%.  There it is also clear that the misclassifications (marked with red circles) all occur around two boundaries. This is understandable and typical because of the effect of the finite size of the training data set.
It is important to look into possible nonadiabatic effects, because the time-evolution state cannot exactly follow the track of the time evolving Hamiltonian, so long as the adiabatic protocol is completed with a finite duration.   Fig.~\ref{f2}(c) compares  the output of actual time evolution with the idealized time evolution with perfect adiabaticity.  Their results match each other with a reasonably high accuracy, with the roughness of the actual output over a finite duration somewhat expected due to {nonadiabatic effects associated with finite-time nature of the simulation}. It is also of interest to inspect the difference between the actual time-evolving state $|\phi_{\mathrm{sim}}(t)\rangle $ and the ground state $ \ket{\phi_g(t)}$ of the corresponding instantaneous Hamiltonian.  To that end we introduce the fidelity $ f $, which is $ |\braket{\phi_\mathrm{sim}(t)}{\phi_g(t)}|^2 $ as a function of time.  This fidelity ranges in $ [0, 1] $, with $ f=1 $ if these two state are the same and $ f=0 $ if they are orthogonal. In Fig.~\ref{f2}(d), we select a sample with  $ x=0 $ from the test data and check the time dependence of the defined fidelity over the entire duration of the adiabatic learning protocol.  It is seen there the fidelity would stay close to unity for finite-time operations with $g=20$.  That is, to suppress possible nonadiabatic effects on the accuracy of our classification task, the important adiabatic parameter $g$ should be at the order of $10$ and the state fidelity is expected to be around $99.9\%$, which is within the reach of today's quantum control technologies. 

\subsection{Case II}

Having benchmarked our protocol in a one-dimensional classification example, we next stretch it to a two-dimensional binary classification task to check our protocol's performance further. 

As shown in Fig.~\ref{f3}(a), the distribution is as follows: in a square with width 2, a circle divide it into two parts with the same area. The radius is $ \sqrt{2/\pi} $. Like the situation in the previous case, we set the label of the part (blue) outside this circle as $ y=1 $ and the other part (green) as $ y=0 $. Input is a two-dimensional vector $ [x_1, x_2]$, with $x_1,x_2\in[-1, 1] $ . The target output for label 1 is 1 and label 0 is $-1$. 

In our learning structure, the variational block is the same as that in the previous case.  The encoding block is now comprised of two rotations corresponding to the two elements of input. One is the rotation around the $z$ axis by $ x_1 $, followed by the rotation around the $y$ axis by $ x_2 $. We still repeat this learning unit for three times as an example.  So all the variational parameters make up a $ 3\times 3=9 $-dimensional vector. The initial state is still the ground state $ \ket{0} $ of Hamiltonian $ -\sigma_z $, and the observable for measurement at the end is still chosen to be  $ \sigma_z $.
\begin{figure}[htbp]
	\includegraphics[scale=0.55]{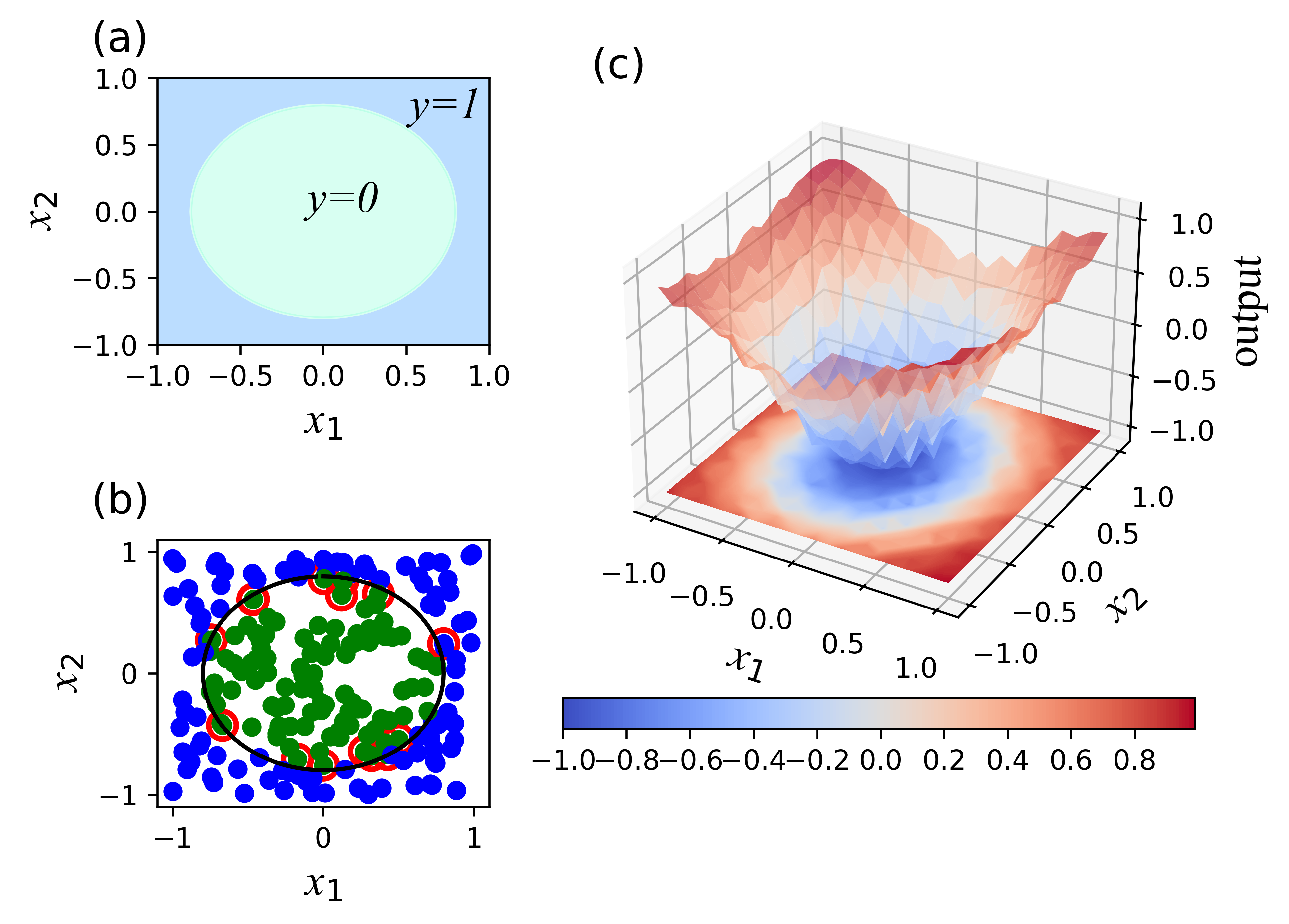}
	\caption{\label{f3}{Two-dimensional binary classification.}  (a) The label distribution inside a square of length 2, with the boundary of the two labels being a circle with a radius $ \sqrt{2/\pi} $. (b) The classification result of 200 randomly selected samples. These wrongly classified samples are marked with red circles and locate around the boundary. The accuracy is $92.5\% $. (c) The output distribution in this square. The surface is somewhat rough, which is again due to nonadiabatic effects in our actual time evolution.  This is also reflected from its projection at the bottom, where the boundary (white part,  output $=0 $) for classification is seen to be not a perfect circle.}
\end{figure}

It is also natural to train our variational parameters after transforming rotation operations on time dependent Hamiltonian to the corresponding quantum circuits. 200 samples are selected randomly as the training data and the loss function is still the L2 norm as we use in the previous case. The optimal parameter vector is [-0.268, 1.628, 0.0176, 2.367, 0.1684, 2.796, 1.044, 1.616, 0.866] with accuracy $ 91.5\% $. 

The adiabatic evolution is simulated again with adiabatic time-scale parameter $g= 0.01/0.0005=20$ per angle change as in previous case. We verify the performance of such an adiabatic learning protocol with a test data set consisting of 200 samples. The obtained result is presented in Fig.~\ref{f3}(b). The points classified wrongly are marked with red circles. The classification accuracy is found to be $ 92.5\% $. Similar to the previous case, these wrong points are located around the boundary (the black circle). Evidently then, the adopted learning structure has effectively learned the pattern of this two-dimensional binary distribution. Finally, the output of adiabatic learning for all the input data is shown in Fig.~\ref{f3}(c). The color represents the value of output. The surface made up of output is not as smooth as it should be theoretically. {It is again due to finite-time nature of our numerical simulations.} From the contour plane projected at the bottom in Fig.~\ref{f3}(c), we can confirm that though the classification boundary is not a perfect circle, it really reflects the correct pattern with an acceptable error. Our results here demonstrate again that adiabatic learning structure advocated above can achieve the same learning capability in comparison to others based on more general quantum circuits.  The gained advantage is then the following: within some acceptable small nonadaiabatic effects, the state of the system always stays at the ground state of the instantaneous Hamiltonian, thus allowing to carry out adiabatic weak measurements at any point of the learning protocol and hence possibly bypassing the annoying repetition of projective measurements in obtaining the expectation value of an observable.

\section{Conclusion}

In this work, we have proposed an alternative quantum learning structure, which allows its ultimate integration with the innovative  weak adiabatic measurement scheme proposed by Aharonov and Vaidman \cite{aharonov1988result}.  Because the output state in conventional quantum learning protocols is usually not the eigenstate of a pre-chosen observable, the repetition of many measurements, which is the strong projective measurement by default, always destroys the final state and hence may not fully use the information of the final state from a long sequence of quantum operations.  The key message from this paper is that the quantum learning structure can be integrated with weak adiabatic measurement schemes by always letting the time-evolving states be one of its energy states. To that end, instead of encoding the learning process into unitary operators, the rotation on the Hamiltonian matrix along a set of axes is used to construct a learning structure.  We show that our learning protocol should have the same learning capacity when compared with the traditional methods based on more general quantum circuits.  Theoretically, it is now possible to imagine that at the end of quantum operations, the expectation value of an observable can be obtained with a single measurement with a reasonable good precision, without collapsing the state under measurement. This will greatly facilitate the optimization process in variational learning by reducing the complexity of output measurement on a quantum learning machine.  It should be even more useful in high-dimensional systems where eigenvalues of an observable are likely to have a lager variance. {In addition, our learning protocol allows us to probe the intermediate state without disturbing the computation or learning process. Unlike a conventional computation processes regarded as a black box, the flow of our protocol thus becomes transparent and any unexpected error can be monitored along the way without destroying the time evolving state.  Actual performance of our protocol in real experimental platforms should be explored in the near future.

Recently, some other learning structures are being developed, such as deep multi-layer perceptrons \cite{alam2022deepqmlp} and quantum-classical convolutional neural network(CNN) \cite{henderson2020quanvolutional,liu2021hybrid}.  We note that more layers needed would require more measurements in the interior of their learning processes. How to reduce complications in the measurement steps there should be urgent and our adiabatic learning protocol likely brings a long-term solution when these learning structure can be implemented by adiabatic time evolution and then integrated with adiabatic measurement schemes.  
\

\appendix
\section{The equivalence between an adiabatic evolution and a quantum circuit}

As explained in the method part, {a traceless Hamiltonian in a $D$-dimensional Hilbert space can be decomposed with $D^2-1 $ traceless elements $e_i, i\in \{1,\dots,D^2-1\}$. It means that a $ D^2-1 $ dimensional unit vector $ \vec{n} $ represents a Hamiltonian in tersm of $ H=\vec{n}\cdot\vec{e} $. $C_{abc}$ is the structure constant with $[e_a,e_b]=2i\sum_{c=1}^{D^2-1}C_{abc}e_c$. After a selected rotational operation $ R $ around a $D^2-1$ dimensional unit vector $\vec{m}$ by angle $\theta$, the new unit vector $ \vec{n'}=R(\vec{n})$ representing a new Hamiltonian $ H' $ can be regarded as the result of a unitary operator $ U $ acting on $ H $, $ H'= UHU^\dagger $.  

Here we give the explicit mapping relation to find the corresponding unitary operator $U$. As shown in the main text, the selected rotational operation around $ \vec{m} $ by $\theta$ is in the form of $ R=e^{\theta A} $ with $ A_{ab}=\sum_{c}C_{abc}m_c $.  As the rotation angle approaches 0, i.e., $ \theta \to 0 $,  a straightforward relation emerges:
\begin{equation}
	\begin{split}
		n'_a&=\sum_{b}R_{ab}n_b\\
			&=\sum_{b}(\delta_{ab}+\theta\sum_{c}m_cC_{abc}+ \\
			& \quad \frac{1}{2}\theta ^2\sum_{cde}C_{acd}C_{cbe}m_dm_e+\dots)n_b			
	\end{split}
\end{equation}
The effect on the Hamiltonian then becomes the following: 
\begin{widetext}
\begin{equation}
	\begin{split}
		\vec{n'}\cdot\vec{e}&=\vec{n}\cdot\vec{e}+\theta\sum_{abc}m_cn_bC_{abc}e_a+\\& \quad \frac{1}{2}\theta ^2\sum_{abcde}C_{acd}C_{cbe}m_dm_en_be_a+\dots\\
						&=\vec{n}\cdot\vec{e}+i(\theta/2)\sum_{ab}m_an_b[e_a,e_b]+ \frac{1}{2}(i\theta/2)^2\sum_{abcd}2iC_{cbd}[e_a,e_d]+\dots\\
						&=\vec{n}\cdot\vec{e}+i(\theta/2)\sum_{ab}m_an_b[e_a,e_b]+ \frac{1}{2}(i\theta/2)^2\sum_{abc}m_an_bm_c(e_a[e_c,e_b]+[e_b,e_a]e_c)+\dots\\
						&=(I+i(\theta/2)\vec{m}\cdot\vec{e}+\frac{1}{2}i[(\theta/2)\vec{m}\cdot\vec{e}]^2+\dots)\vec{n}\cdot\vec{e}(I-i(\theta/2)\vec{m}\cdot\vec{e}+\frac{1}{2}-i[(\theta/2)\vec{m}\cdot\vec{e}]^2+\dots)\\
						&=e^{i(\theta/2)\vec{m}\cdot\vec{e}}\vec{n}\cdot\vec{e}e^{-i(\theta/2)\vec{m}\cdot\vec{e}}\\
						&=U(\theta)\vec{n}\cdot\vec{e}U^\dagger(\theta).				
	\end{split}
\end{equation}
\end{widetext}
Thus, with an infinitesimal rotation, the relation between a unitary transformation $ U(\theta)=e^{i(\theta/2)\vec{m}\cdot\vec{e}} $ experienced by a Hamiltonian and the corresponding rotational operation $ R=e^{\theta A} $ on the corresponding unit vector is obvious.  With multiplication of many such infinitesimal rotations, it becomes obvious that a corresponding unitary operator can be found for any rotation angle $ \theta $.  Furthermore, any unitary operator in a quantum circuit can be written as $U(\theta)$ in terms of the exponential of some Hermitian operators (apart from a global phase).  Thus, from a unitary operator considered in a quantum circuit, a corresponding rotational operator can always be identified.   This establishes the mapping between a rotational operation in our protocol and unitary operators in a conventional quantum circuit. As argued in the main text,  our protocol is hence expected to have the same expressivity as in a conventional quantum circuit. This also explains why we can first find optimal variational parameters from the corresponding quantum circuits in the training step and then finally do the finite-time rotational operations to implement our adiabatic learning protocol.}


\begin{thebibliography}{43}%
\makeatletter
\providecommand \@ifxundefined [1]{%
 \@ifx{#1\undefined}
}%
\providecommand \@ifnum [1]{%
 \ifnum #1\expandafter \@firstoftwo
 \else \expandafter \@secondoftwo
 \fi
}%
\providecommand \@ifx [1]{%
 \ifx #1\expandafter \@firstoftwo
 \else \expandafter \@secondoftwo
 \fi
}%
\providecommand \natexlab [1]{#1}%
\providecommand \enquote  [1]{``#1''}%
\providecommand \bibnamefont  [1]{#1}%
\providecommand \bibfnamefont [1]{#1}%
\providecommand \citenamefont [1]{#1}%
\providecommand \href@noop [0]{\@secondoftwo}%
\providecommand \href [0]{\begingroup \@sanitize@url \@href}%
\providecommand \@href[1]{\@@startlink{#1}\@@href}%
\providecommand \@@href[1]{\endgroup#1\@@endlink}%
\providecommand \@sanitize@url [0]{\catcode `\\12\catcode `\$12\catcode
  `\&12\catcode `\#12\catcode `\^12\catcode `\_12\catcode `\%12\relax}%
\providecommand \@@startlink[1]{}%
\providecommand \@@endlink[0]{}%
\providecommand \url  [0]{\begingroup\@sanitize@url \@url }%
\providecommand \@url [1]{\endgroup\@href {#1}{\urlprefix }}%
\providecommand \urlprefix  [0]{URL }%
\providecommand \Eprint [0]{\href }%
\providecommand \doibase [0]{https://doi.org/}%
\providecommand \selectlanguage [0]{\@gobble}%
\providecommand \bibinfo  [0]{\@secondoftwo}%
\providecommand \bibfield  [0]{\@secondoftwo}%
\providecommand \translation [1]{[#1]}%
\providecommand \BibitemOpen [0]{}%
\providecommand \bibitemStop [0]{}%
\providecommand \bibitemNoStop [0]{.\EOS\space}%
\providecommand \EOS [0]{\spacefactor3000\relax}%
\providecommand \BibitemShut  [1]{\csname bibitem#1\endcsname}%
\let\auto@bib@innerbib\@empty
\bibitem [{\citenamefont {Lloyd}(1996)}]{lloyd1996universal}%
  \BibitemOpen
  \bibfield  {author} {\bibinfo {author} {\bibfnamefont {S.}~\bibnamefont
  {Lloyd}},\ }\bibfield  {title} {\bibinfo {title} {Universal quantum
  simulators},\ }\href@noop {} {\bibfield  {journal} {\bibinfo  {journal}
  {Science}\ }\textbf {\bibinfo {volume} {273}},\ \bibinfo {pages} {1073}
  (\bibinfo {year} {1996})}\BibitemShut {NoStop}%
\bibitem [{\citenamefont {DiVincenzo}(1995)}]{divincenzo1995quantum}%
  \BibitemOpen
  \bibfield  {author} {\bibinfo {author} {\bibfnamefont {D.~P.}\ \bibnamefont
  {DiVincenzo}},\ }\bibfield  {title} {\bibinfo {title} {Quantum computation},\
  }\href@noop {} {\bibfield  {journal} {\bibinfo  {journal} {Science}\ }\textbf
  {\bibinfo {volume} {270}},\ \bibinfo {pages} {255} (\bibinfo {year}
  {1995})}\BibitemShut {NoStop}%
\bibitem [{\citenamefont {Nielsen}\ and\ \citenamefont
  {Chuang}(2002)}]{nielsen2002quantum}%
  \BibitemOpen
  \bibfield  {author} {\bibinfo {author} {\bibfnamefont {M.~A.}\ \bibnamefont
  {Nielsen}}\ and\ \bibinfo {author} {\bibfnamefont {I.}~\bibnamefont
  {Chuang}},\ }\href@noop {} {\bibinfo {title} {Quantum computation and quantum
  information}} (\bibinfo {year} {2002})\BibitemShut {NoStop}%
\bibitem [{\citenamefont {Shor}(1999)}]{shor1999polynomial}%
  \BibitemOpen
  \bibfield  {author} {\bibinfo {author} {\bibfnamefont {P.~W.}\ \bibnamefont
  {Shor}},\ }\bibfield  {title} {\bibinfo {title} {Polynomial-time algorithms
  for prime factorization and discrete logarithms on a quantum computer},\
  }\href@noop {} {\bibfield  {journal} {\bibinfo  {journal} {SIAM review}\
  }\textbf {\bibinfo {volume} {41}},\ \bibinfo {pages} {303} (\bibinfo {year}
  {1999})}\BibitemShut {NoStop}%
\bibitem [{\citenamefont {Grover}(1996)}]{grover1996fast}%
  \BibitemOpen
  \bibfield  {author} {\bibinfo {author} {\bibfnamefont {L.~K.}\ \bibnamefont
  {Grover}},\ }\bibfield  {title} {\bibinfo {title} {A fast quantum mechanical
  algorithm for database search},\ }in\ \href@noop {} {\emph {\bibinfo
  {booktitle} {Proceedings of the twenty-eighth annual ACM symposium on Theory
  of computing}}}\ (\bibinfo {year} {1996})\ pp.\ \bibinfo {pages}
  {212--219}\BibitemShut {NoStop}%
\bibitem [{\citenamefont {Montanaro}(2016)}]{montanaro2016quantum}%
  \BibitemOpen
  \bibfield  {author} {\bibinfo {author} {\bibfnamefont {A.}~\bibnamefont
  {Montanaro}},\ }\bibfield  {title} {\bibinfo {title} {Quantum algorithms: an
  overview},\ }\href@noop {} {\bibfield  {journal} {\bibinfo  {journal} {npj
  Quantum Information}\ }\textbf {\bibinfo {volume} {2}},\ \bibinfo {pages} {1}
  (\bibinfo {year} {2016})}\BibitemShut {NoStop}%
\bibitem [{\citenamefont {Liu}\ \emph {et~al.}(2021{\natexlab{a}})\citenamefont
  {Liu}, \citenamefont {Arunachalam},\ and\ \citenamefont
  {Temme}}]{liu2021rigorous}%
  \BibitemOpen
  \bibfield  {author} {\bibinfo {author} {\bibfnamefont {Y.}~\bibnamefont
  {Liu}}, \bibinfo {author} {\bibfnamefont {S.}~\bibnamefont {Arunachalam}},\
  and\ \bibinfo {author} {\bibfnamefont {K.}~\bibnamefont {Temme}},\ }\bibfield
   {title} {\bibinfo {title} {A rigorous and robust quantum speed-up in
  supervised machine learning},\ }\href@noop {} {\bibfield  {journal} {\bibinfo
   {journal} {Nature Physics}\ }\textbf {\bibinfo {volume} {17}},\ \bibinfo
  {pages} {1013} (\bibinfo {year} {2021}{\natexlab{a}})}\BibitemShut {NoStop}%
\bibitem [{\citenamefont {Peruzzo}\ \emph {et~al.}(2014)\citenamefont
  {Peruzzo}, \citenamefont {McClean}, \citenamefont {Shadbolt}, \citenamefont
  {Yung}, \citenamefont {Zhou}, \citenamefont {Love}, \citenamefont
  {Aspuru-Guzik},\ and\ \citenamefont {O’brien}}]{peruzzo2014variational}%
  \BibitemOpen
  \bibfield  {author} {\bibinfo {author} {\bibfnamefont {A.}~\bibnamefont
  {Peruzzo}}, \bibinfo {author} {\bibfnamefont {J.}~\bibnamefont {McClean}},
  \bibinfo {author} {\bibfnamefont {P.}~\bibnamefont {Shadbolt}}, \bibinfo
  {author} {\bibfnamefont {M.-H.}\ \bibnamefont {Yung}}, \bibinfo {author}
  {\bibfnamefont {X.-Q.}\ \bibnamefont {Zhou}}, \bibinfo {author}
  {\bibfnamefont {P.~J.}\ \bibnamefont {Love}}, \bibinfo {author}
  {\bibfnamefont {A.}~\bibnamefont {Aspuru-Guzik}},\ and\ \bibinfo {author}
  {\bibfnamefont {J.~L.}\ \bibnamefont {O’brien}},\ }\bibfield  {title}
  {\bibinfo {title} {A variational eigenvalue solver on a photonic quantum
  processor},\ }\href@noop {} {\bibfield  {journal} {\bibinfo  {journal}
  {Nature communications}\ }\textbf {\bibinfo {volume} {5}},\ \bibinfo {pages}
  {1} (\bibinfo {year} {2014})}\BibitemShut {NoStop}%
\bibitem [{\citenamefont {Kandala}\ \emph {et~al.}(2017)\citenamefont
  {Kandala}, \citenamefont {Mezzacapo}, \citenamefont {Temme}, \citenamefont
  {Takita}, \citenamefont {Brink}, \citenamefont {Chow},\ and\ \citenamefont
  {Gambetta}}]{kandala2017hardware}%
  \BibitemOpen
  \bibfield  {author} {\bibinfo {author} {\bibfnamefont {A.}~\bibnamefont
  {Kandala}}, \bibinfo {author} {\bibfnamefont {A.}~\bibnamefont {Mezzacapo}},
  \bibinfo {author} {\bibfnamefont {K.}~\bibnamefont {Temme}}, \bibinfo
  {author} {\bibfnamefont {M.}~\bibnamefont {Takita}}, \bibinfo {author}
  {\bibfnamefont {M.}~\bibnamefont {Brink}}, \bibinfo {author} {\bibfnamefont
  {J.~M.}\ \bibnamefont {Chow}},\ and\ \bibinfo {author} {\bibfnamefont
  {J.~M.}\ \bibnamefont {Gambetta}},\ }\bibfield  {title} {\bibinfo {title}
  {Hardware-efficient variational quantum eigensolver for small molecules and
  quantum magnets},\ }\href@noop {} {\bibfield  {journal} {\bibinfo  {journal}
  {Nature}\ }\textbf {\bibinfo {volume} {549}},\ \bibinfo {pages} {242}
  (\bibinfo {year} {2017})}\BibitemShut {NoStop}%
\bibitem [{\citenamefont {Vandersypen}\ \emph {et~al.}(2001)\citenamefont
  {Vandersypen}, \citenamefont {Steffen}, \citenamefont {Breyta}, \citenamefont
  {Yannoni}, \citenamefont {Sherwood},\ and\ \citenamefont
  {Chuang}}]{vandersypen2001experimental}%
  \BibitemOpen
  \bibfield  {author} {\bibinfo {author} {\bibfnamefont {L.~M.}\ \bibnamefont
  {Vandersypen}}, \bibinfo {author} {\bibfnamefont {M.}~\bibnamefont
  {Steffen}}, \bibinfo {author} {\bibfnamefont {G.}~\bibnamefont {Breyta}},
  \bibinfo {author} {\bibfnamefont {C.~S.}\ \bibnamefont {Yannoni}}, \bibinfo
  {author} {\bibfnamefont {M.~H.}\ \bibnamefont {Sherwood}},\ and\ \bibinfo
  {author} {\bibfnamefont {I.~L.}\ \bibnamefont {Chuang}},\ }\bibfield  {title}
  {\bibinfo {title} {Experimental realization of shor's quantum factoring
  algorithm using nuclear magnetic resonance},\ }\href@noop {} {\bibfield
  {journal} {\bibinfo  {journal} {Nature}\ }\textbf {\bibinfo {volume} {414}},\
  \bibinfo {pages} {883} (\bibinfo {year} {2001})}\BibitemShut {NoStop}%
\bibitem [{\citenamefont {Arute}\ \emph {et~al.}(2019)\citenamefont {Arute},
  \citenamefont {Arya}, \citenamefont {Babbush}, \citenamefont {Bacon},
  \citenamefont {Bardin}, \citenamefont {Barends}, \citenamefont {Biswas},
  \citenamefont {Boixo}, \citenamefont {Brandao}, \citenamefont {Buell} \emph
  {et~al.}}]{arute2019quantum}%
  \BibitemOpen
  \bibfield  {author} {\bibinfo {author} {\bibfnamefont {F.}~\bibnamefont
  {Arute}}, \bibinfo {author} {\bibfnamefont {K.}~\bibnamefont {Arya}},
  \bibinfo {author} {\bibfnamefont {R.}~\bibnamefont {Babbush}}, \bibinfo
  {author} {\bibfnamefont {D.}~\bibnamefont {Bacon}}, \bibinfo {author}
  {\bibfnamefont {J.~C.}\ \bibnamefont {Bardin}}, \bibinfo {author}
  {\bibfnamefont {R.}~\bibnamefont {Barends}}, \bibinfo {author} {\bibfnamefont
  {R.}~\bibnamefont {Biswas}}, \bibinfo {author} {\bibfnamefont
  {S.}~\bibnamefont {Boixo}}, \bibinfo {author} {\bibfnamefont {F.~G.}\
  \bibnamefont {Brandao}}, \bibinfo {author} {\bibfnamefont {D.~A.}\
  \bibnamefont {Buell}}, \emph {et~al.},\ }\bibfield  {title} {\bibinfo {title}
  {Quantum supremacy using a programmable superconducting processor},\
  }\href@noop {} {\bibfield  {journal} {\bibinfo  {journal} {Nature}\ }\textbf
  {\bibinfo {volume} {574}},\ \bibinfo {pages} {505} (\bibinfo {year}
  {2019})}\BibitemShut {NoStop}%
\bibitem [{\citenamefont {Carleo}\ \emph {et~al.}(2019)\citenamefont {Carleo},
  \citenamefont {Cirac}, \citenamefont {Cranmer}, \citenamefont {Daudet},
  \citenamefont {Schuld}, \citenamefont {Tishby}, \citenamefont
  {Vogt-Maranto},\ and\ \citenamefont {Zdeborov{\'a}}}]{carleo2019machine}%
  \BibitemOpen
  \bibfield  {author} {\bibinfo {author} {\bibfnamefont {G.}~\bibnamefont
  {Carleo}}, \bibinfo {author} {\bibfnamefont {I.}~\bibnamefont {Cirac}},
  \bibinfo {author} {\bibfnamefont {K.}~\bibnamefont {Cranmer}}, \bibinfo
  {author} {\bibfnamefont {L.}~\bibnamefont {Daudet}}, \bibinfo {author}
  {\bibfnamefont {M.}~\bibnamefont {Schuld}}, \bibinfo {author} {\bibfnamefont
  {N.}~\bibnamefont {Tishby}}, \bibinfo {author} {\bibfnamefont
  {L.}~\bibnamefont {Vogt-Maranto}},\ and\ \bibinfo {author} {\bibfnamefont
  {L.}~\bibnamefont {Zdeborov{\'a}}},\ }\bibfield  {title} {\bibinfo {title}
  {Machine learning and the physical sciences},\ }\href@noop {} {\bibfield
  {journal} {\bibinfo  {journal} {Reviews of Modern Physics}\ }\textbf
  {\bibinfo {volume} {91}},\ \bibinfo {pages} {045002} (\bibinfo {year}
  {2019})}\BibitemShut {NoStop}%
\bibitem [{\citenamefont {Biamonte}\ \emph {et~al.}(2017)\citenamefont
  {Biamonte}, \citenamefont {Wittek}, \citenamefont {Pancotti}, \citenamefont
  {Rebentrost}, \citenamefont {Wiebe},\ and\ \citenamefont
  {Lloyd}}]{biamonte2017quantum}%
  \BibitemOpen
  \bibfield  {author} {\bibinfo {author} {\bibfnamefont {J.}~\bibnamefont
  {Biamonte}}, \bibinfo {author} {\bibfnamefont {P.}~\bibnamefont {Wittek}},
  \bibinfo {author} {\bibfnamefont {N.}~\bibnamefont {Pancotti}}, \bibinfo
  {author} {\bibfnamefont {P.}~\bibnamefont {Rebentrost}}, \bibinfo {author}
  {\bibfnamefont {N.}~\bibnamefont {Wiebe}},\ and\ \bibinfo {author}
  {\bibfnamefont {S.}~\bibnamefont {Lloyd}},\ }\bibfield  {title} {\bibinfo
  {title} {Quantum machine learning},\ }\href@noop {} {\bibfield  {journal}
  {\bibinfo  {journal} {Nature}\ }\textbf {\bibinfo {volume} {549}},\ \bibinfo
  {pages} {195} (\bibinfo {year} {2017})}\BibitemShut {NoStop}%
\bibitem [{\citenamefont {Cong}\ \emph {et~al.}(2019)\citenamefont {Cong},
  \citenamefont {Choi},\ and\ \citenamefont {Lukin}}]{cong2019quantum}%
  \BibitemOpen
  \bibfield  {author} {\bibinfo {author} {\bibfnamefont {I.}~\bibnamefont
  {Cong}}, \bibinfo {author} {\bibfnamefont {S.}~\bibnamefont {Choi}},\ and\
  \bibinfo {author} {\bibfnamefont {M.~D.}\ \bibnamefont {Lukin}},\ }\bibfield
  {title} {\bibinfo {title} {Quantum convolutional neural networks},\
  }\href@noop {} {\bibfield  {journal} {\bibinfo  {journal} {Nature Physics}\
  }\textbf {\bibinfo {volume} {15}},\ \bibinfo {pages} {1273} (\bibinfo {year}
  {2019})}\BibitemShut {NoStop}%
\bibitem [{\citenamefont {Schuld}\ and\ \citenamefont
  {Killoran}(2019)}]{schuld2019quantum}%
  \BibitemOpen
  \bibfield  {author} {\bibinfo {author} {\bibfnamefont {M.}~\bibnamefont
  {Schuld}}\ and\ \bibinfo {author} {\bibfnamefont {N.}~\bibnamefont
  {Killoran}},\ }\bibfield  {title} {\bibinfo {title} {Quantum machine learning
  in feature hilbert spaces},\ }\href@noop {} {\bibfield  {journal} {\bibinfo
  {journal} {Physical review letters}\ }\textbf {\bibinfo {volume} {122}},\
  \bibinfo {pages} {040504} (\bibinfo {year} {2019})}\BibitemShut {NoStop}%
\bibitem [{\citenamefont {McClean}\ \emph {et~al.}(2016)\citenamefont
  {McClean}, \citenamefont {Romero}, \citenamefont {Babbush},\ and\
  \citenamefont {Aspuru-Guzik}}]{mcclean2016theory}%
  \BibitemOpen
  \bibfield  {author} {\bibinfo {author} {\bibfnamefont {J.~R.}\ \bibnamefont
  {McClean}}, \bibinfo {author} {\bibfnamefont {J.}~\bibnamefont {Romero}},
  \bibinfo {author} {\bibfnamefont {R.}~\bibnamefont {Babbush}},\ and\ \bibinfo
  {author} {\bibfnamefont {A.}~\bibnamefont {Aspuru-Guzik}},\ }\bibfield
  {title} {\bibinfo {title} {The theory of variational hybrid quantum-classical
  algorithms},\ }\href@noop {} {\bibfield  {journal} {\bibinfo  {journal} {New
  Journal of Physics}\ }\textbf {\bibinfo {volume} {18}},\ \bibinfo {pages}
  {023023} (\bibinfo {year} {2016})}\BibitemShut {NoStop}%
\bibitem [{\citenamefont {Yuan}\ \emph {et~al.}(2019)\citenamefont {Yuan},
  \citenamefont {Endo}, \citenamefont {Zhao}, \citenamefont {Li},\ and\
  \citenamefont {Benjamin}}]{yuan2019theory}%
  \BibitemOpen
  \bibfield  {author} {\bibinfo {author} {\bibfnamefont {X.}~\bibnamefont
  {Yuan}}, \bibinfo {author} {\bibfnamefont {S.}~\bibnamefont {Endo}}, \bibinfo
  {author} {\bibfnamefont {Q.}~\bibnamefont {Zhao}}, \bibinfo {author}
  {\bibfnamefont {Y.}~\bibnamefont {Li}},\ and\ \bibinfo {author}
  {\bibfnamefont {S.~C.}\ \bibnamefont {Benjamin}},\ }\bibfield  {title}
  {\bibinfo {title} {Theory of variational quantum simulation},\ }\href@noop {}
  {\bibfield  {journal} {\bibinfo  {journal} {Quantum}\ }\textbf {\bibinfo
  {volume} {3}},\ \bibinfo {pages} {191} (\bibinfo {year} {2019})}\BibitemShut
  {NoStop}%
\bibitem [{\citenamefont {Goto}\ \emph {et~al.}(2021)\citenamefont {Goto},
  \citenamefont {Tran},\ and\ \citenamefont {Nakajima}}]{goto2021universal}%
  \BibitemOpen
  \bibfield  {author} {\bibinfo {author} {\bibfnamefont {T.}~\bibnamefont
  {Goto}}, \bibinfo {author} {\bibfnamefont {Q.~H.}\ \bibnamefont {Tran}},\
  and\ \bibinfo {author} {\bibfnamefont {K.}~\bibnamefont {Nakajima}},\
  }\bibfield  {title} {\bibinfo {title} {Universal approximation property of
  quantum machine learning models in quantum-enhanced feature spaces},\
  }\href@noop {} {\bibfield  {journal} {\bibinfo  {journal} {Physical Review
  Letters}\ }\textbf {\bibinfo {volume} {127}},\ \bibinfo {pages} {090506}
  (\bibinfo {year} {2021})}\BibitemShut {NoStop}%
\bibitem [{\citenamefont {Mitarai}\ \emph {et~al.}(2018)\citenamefont
  {Mitarai}, \citenamefont {Negoro}, \citenamefont {Kitagawa},\ and\
  \citenamefont {Fujii}}]{mitarai2018quantum}%
  \BibitemOpen
  \bibfield  {author} {\bibinfo {author} {\bibfnamefont {K.}~\bibnamefont
  {Mitarai}}, \bibinfo {author} {\bibfnamefont {M.}~\bibnamefont {Negoro}},
  \bibinfo {author} {\bibfnamefont {M.}~\bibnamefont {Kitagawa}},\ and\
  \bibinfo {author} {\bibfnamefont {K.}~\bibnamefont {Fujii}},\ }\bibfield
  {title} {\bibinfo {title} {Quantum circuit learning},\ }\href@noop {}
  {\bibfield  {journal} {\bibinfo  {journal} {Physical Review A}\ }\textbf
  {\bibinfo {volume} {98}},\ \bibinfo {pages} {032309} (\bibinfo {year}
  {2018})}\BibitemShut {NoStop}%
\bibitem [{\citenamefont {Havl{\'\i}{\v{c}}ek}\ \emph
  {et~al.}(2019)\citenamefont {Havl{\'\i}{\v{c}}ek}, \citenamefont
  {C{\'o}rcoles}, \citenamefont {Temme}, \citenamefont {Harrow}, \citenamefont
  {Kandala}, \citenamefont {Chow},\ and\ \citenamefont
  {Gambetta}}]{havlivcek2019supervised}%
  \BibitemOpen
  \bibfield  {author} {\bibinfo {author} {\bibfnamefont {V.}~\bibnamefont
  {Havl{\'\i}{\v{c}}ek}}, \bibinfo {author} {\bibfnamefont {A.~D.}\
  \bibnamefont {C{\'o}rcoles}}, \bibinfo {author} {\bibfnamefont
  {K.}~\bibnamefont {Temme}}, \bibinfo {author} {\bibfnamefont {A.~W.}\
  \bibnamefont {Harrow}}, \bibinfo {author} {\bibfnamefont {A.}~\bibnamefont
  {Kandala}}, \bibinfo {author} {\bibfnamefont {J.~M.}\ \bibnamefont {Chow}},\
  and\ \bibinfo {author} {\bibfnamefont {J.~M.}\ \bibnamefont {Gambetta}},\
  }\bibfield  {title} {\bibinfo {title} {Supervised learning with
  quantum-enhanced feature spaces},\ }\href@noop {} {\bibfield  {journal}
  {\bibinfo  {journal} {Nature}\ }\textbf {\bibinfo {volume} {567}},\ \bibinfo
  {pages} {209} (\bibinfo {year} {2019})}\BibitemShut {NoStop}%
\bibitem [{\citenamefont {Lloyd}\ \emph {et~al.}(2020)\citenamefont {Lloyd},
  \citenamefont {Schuld}, \citenamefont {Ijaz}, \citenamefont {Izaac},\ and\
  \citenamefont {Killoran}}]{lloyd2020quantum}%
  \BibitemOpen
  \bibfield  {author} {\bibinfo {author} {\bibfnamefont {S.}~\bibnamefont
  {Lloyd}}, \bibinfo {author} {\bibfnamefont {M.}~\bibnamefont {Schuld}},
  \bibinfo {author} {\bibfnamefont {A.}~\bibnamefont {Ijaz}}, \bibinfo {author}
  {\bibfnamefont {J.}~\bibnamefont {Izaac}},\ and\ \bibinfo {author}
  {\bibfnamefont {N.}~\bibnamefont {Killoran}},\ }\bibfield  {title} {\bibinfo
  {title} {Quantum embeddings for machine learning},\ }\href@noop {} {\bibfield
   {journal} {\bibinfo  {journal} {arXiv preprint arXiv:2001.03622}\ }
  (\bibinfo {year} {2020})}\BibitemShut {NoStop}%
\bibitem [{\citenamefont {Schuld}\ \emph {et~al.}(2020)\citenamefont {Schuld},
  \citenamefont {Bocharov}, \citenamefont {Svore},\ and\ \citenamefont
  {Wiebe}}]{schuld2020circuit}%
  \BibitemOpen
  \bibfield  {author} {\bibinfo {author} {\bibfnamefont {M.}~\bibnamefont
  {Schuld}}, \bibinfo {author} {\bibfnamefont {A.}~\bibnamefont {Bocharov}},
  \bibinfo {author} {\bibfnamefont {K.~M.}\ \bibnamefont {Svore}},\ and\
  \bibinfo {author} {\bibfnamefont {N.}~\bibnamefont {Wiebe}},\ }\bibfield
  {title} {\bibinfo {title} {Circuit-centric quantum classifiers},\ }\href@noop
  {} {\bibfield  {journal} {\bibinfo  {journal} {Physical Review A}\ }\textbf
  {\bibinfo {volume} {101}},\ \bibinfo {pages} {032308} (\bibinfo {year}
  {2020})}\BibitemShut {NoStop}%
\bibitem [{\citenamefont {P{\'e}rez-Salinas}\ \emph {et~al.}(2021)\citenamefont
  {P{\'e}rez-Salinas}, \citenamefont {L{\'o}pez-N{\'u}{\~n}ez}, \citenamefont
  {Garc{\'\i}a-S{\'a}ez}, \citenamefont {Forn-D{\'\i}az},\ and\ \citenamefont
  {Latorre}}]{perez2021one}%
  \BibitemOpen
  \bibfield  {author} {\bibinfo {author} {\bibfnamefont {A.}~\bibnamefont
  {P{\'e}rez-Salinas}}, \bibinfo {author} {\bibfnamefont {D.}~\bibnamefont
  {L{\'o}pez-N{\'u}{\~n}ez}}, \bibinfo {author} {\bibfnamefont
  {A.}~\bibnamefont {Garc{\'\i}a-S{\'a}ez}}, \bibinfo {author} {\bibfnamefont
  {P.}~\bibnamefont {Forn-D{\'\i}az}},\ and\ \bibinfo {author} {\bibfnamefont
  {J.~I.}\ \bibnamefont {Latorre}},\ }\bibfield  {title} {\bibinfo {title} {One
  qubit as a universal approximant},\ }\href@noop {} {\bibfield  {journal}
  {\bibinfo  {journal} {Physical Review A}\ }\textbf {\bibinfo {volume}
  {104}},\ \bibinfo {pages} {012405} (\bibinfo {year} {2021})}\BibitemShut
  {NoStop}%
\bibitem [{\citenamefont {Von~Neumann}(2018)}]{von2018mathematical}%
  \BibitemOpen
  \bibfield  {author} {\bibinfo {author} {\bibfnamefont {J.}~\bibnamefont
  {Von~Neumann}},\ }\bibfield  {title} {\bibinfo {title} {Mathematical
  foundations of quantum mechanics},\ }in\ \href@noop {} {\emph {\bibinfo
  {booktitle} {Mathematical Foundations of Quantum Mechanics}}}\ (\bibinfo
  {publisher} {Princeton university press},\ \bibinfo {year}
  {2018})\BibitemShut {NoStop}%
\bibitem [{\citenamefont {Alonso}\ and\ \citenamefont
  {Garc{\'\i}a}(2022)}]{alonso2022single}%
  \BibitemOpen
  \bibfield  {author} {\bibinfo {author} {\bibfnamefont {D.}~\bibnamefont
  {Alonso}}\ and\ \bibinfo {author} {\bibfnamefont {A.~R.}\ \bibnamefont
  {Garc{\'\i}a}},\ }\bibfield  {title} {\bibinfo {title} {Single energy
  measurement integral fluctuation theorem and non-projective measurements},\
  }\href@noop {} {\bibfield  {journal} {\bibinfo  {journal} {arXiv preprint
  arXiv:2212.13225}\ } (\bibinfo {year} {2022})}\BibitemShut {NoStop}%
\bibitem [{\citenamefont {Banchi}\ and\ \citenamefont
  {Crooks}(2021)}]{banchi2021measuring}%
  \BibitemOpen
  \bibfield  {author} {\bibinfo {author} {\bibfnamefont {L.}~\bibnamefont
  {Banchi}}\ and\ \bibinfo {author} {\bibfnamefont {G.~E.}\ \bibnamefont
  {Crooks}},\ }\bibfield  {title} {\bibinfo {title} {Measuring analytic
  gradients of general quantum evolution with the stochastic parameter shift
  rule},\ }\href@noop {} {\bibfield  {journal} {\bibinfo  {journal} {Quantum}\
  }\textbf {\bibinfo {volume} {5}},\ \bibinfo {pages} {386} (\bibinfo {year}
  {2021})}\BibitemShut {NoStop}%
\bibitem [{\citenamefont {Huang}\ \emph {et~al.}(2020)\citenamefont {Huang},
  \citenamefont {Kueng},\ and\ \citenamefont {Preskill}}]{huang2020predicting}%
  \BibitemOpen
  \bibfield  {author} {\bibinfo {author} {\bibfnamefont {H.-Y.}\ \bibnamefont
  {Huang}}, \bibinfo {author} {\bibfnamefont {R.}~\bibnamefont {Kueng}},\ and\
  \bibinfo {author} {\bibfnamefont {J.}~\bibnamefont {Preskill}},\ }\bibfield
  {title} {\bibinfo {title} {Predicting many properties of a quantum system
  from very few measurements},\ }\href@noop {} {\bibfield  {journal} {\bibinfo
  {journal} {Nature Physics}\ }\textbf {\bibinfo {volume} {16}},\ \bibinfo
  {pages} {1050} (\bibinfo {year} {2020})}\BibitemShut {NoStop}%
\bibitem [{\citenamefont {Gresch}\ and\ \citenamefont
  {Kliesch}(2023)}]{gresch2023guaranteed}%
  \BibitemOpen
  \bibfield  {author} {\bibinfo {author} {\bibfnamefont {A.}~\bibnamefont
  {Gresch}}\ and\ \bibinfo {author} {\bibfnamefont {M.}~\bibnamefont
  {Kliesch}},\ }\bibfield  {title} {\bibinfo {title} {Guaranteed efficient
  energy estimation of quantum many-body hamiltonians using shadowgrouping},\
  }\href@noop {} {\bibfield  {journal} {\bibinfo  {journal} {arXiv preprint
  arXiv:2301.03385}\ } (\bibinfo {year} {2023})}\BibitemShut {NoStop}%
\bibitem [{\citenamefont {Kohda}\ \emph {et~al.}(2022)\citenamefont {Kohda},
  \citenamefont {Imai}, \citenamefont {Kanno}, \citenamefont {Mitarai},
  \citenamefont {Mizukami},\ and\ \citenamefont {Nakagawa}}]{kohda2022quantum}%
  \BibitemOpen
  \bibfield  {author} {\bibinfo {author} {\bibfnamefont {M.}~\bibnamefont
  {Kohda}}, \bibinfo {author} {\bibfnamefont {R.}~\bibnamefont {Imai}},
  \bibinfo {author} {\bibfnamefont {K.}~\bibnamefont {Kanno}}, \bibinfo
  {author} {\bibfnamefont {K.}~\bibnamefont {Mitarai}}, \bibinfo {author}
  {\bibfnamefont {W.}~\bibnamefont {Mizukami}},\ and\ \bibinfo {author}
  {\bibfnamefont {Y.~O.}\ \bibnamefont {Nakagawa}},\ }\bibfield  {title}
  {\bibinfo {title} {Quantum expectation-value estimation by computational
  basis sampling},\ }\href@noop {} {\bibfield  {journal} {\bibinfo  {journal}
  {Physical Review Research}\ }\textbf {\bibinfo {volume} {4}},\ \bibinfo
  {pages} {033173} (\bibinfo {year} {2022})}\BibitemShut {NoStop}%
\bibitem [{\citenamefont {Solinas}\ \emph {et~al.}(2023)\citenamefont
  {Solinas}, \citenamefont {Caletti},\ and\ \citenamefont
  {Minuto}}]{solinas2023quantum}%
  \BibitemOpen
  \bibfield  {author} {\bibinfo {author} {\bibfnamefont {P.}~\bibnamefont
  {Solinas}}, \bibinfo {author} {\bibfnamefont {S.}~\bibnamefont {Caletti}},\
  and\ \bibinfo {author} {\bibfnamefont {G.}~\bibnamefont {Minuto}},\
  }\bibfield  {title} {\bibinfo {title} {Quantum gradient evaluation through
  quantum non-demolition measurements},\ }\href@noop {} {\bibfield  {journal}
  {\bibinfo  {journal} {arXiv preprint arXiv:2301.07128}\ } (\bibinfo {year}
  {2023})}\BibitemShut {NoStop}%
\bibitem [{\citenamefont {Aharonov}\ \emph {et~al.}(1988)\citenamefont
  {Aharonov}, \citenamefont {Albert},\ and\ \citenamefont
  {Vaidman}}]{aharonov1988result}%
  \BibitemOpen
  \bibfield  {author} {\bibinfo {author} {\bibfnamefont {Y.}~\bibnamefont
  {Aharonov}}, \bibinfo {author} {\bibfnamefont {D.~Z.}\ \bibnamefont
  {Albert}},\ and\ \bibinfo {author} {\bibfnamefont {L.}~\bibnamefont
  {Vaidman}},\ }\bibfield  {title} {\bibinfo {title} {How the result of a
  measurement of a component of the spin of a spin-1/2 particle can turn out to
  be 100},\ }\href@noop {} {\bibfield  {journal} {\bibinfo  {journal} {Physical
  review letters}\ }\textbf {\bibinfo {volume} {60}},\ \bibinfo {pages} {1351}
  (\bibinfo {year} {1988})}\BibitemShut {NoStop}%
\bibitem [{\citenamefont {Aharonov}\ and\ \citenamefont
  {Vaidman}(1993)}]{AHARONOV199338}%
  \BibitemOpen
  \bibfield  {author} {\bibinfo {author} {\bibfnamefont {Y.}~\bibnamefont
  {Aharonov}}\ and\ \bibinfo {author} {\bibfnamefont {L.}~\bibnamefont
  {Vaidman}},\ }\bibfield  {title} {\bibinfo {title} {Measurement of the
  schrödinger wave of a single particle},\ }\href
  {https://doi.org/https://doi.org/10.1016/0375-9601(93)90724-E} {\bibfield
  {journal} {\bibinfo  {journal} {Physics Letters A}\ }\textbf {\bibinfo
  {volume} {178}},\ \bibinfo {pages} {38} (\bibinfo {year} {1993})}\BibitemShut
  {NoStop}%
\bibitem [{\citenamefont {Born}\ and\ \citenamefont
  {Fock}(1928)}]{born1928beweis}%
  \BibitemOpen
  \bibfield  {author} {\bibinfo {author} {\bibfnamefont {M.}~\bibnamefont
  {Born}}\ and\ \bibinfo {author} {\bibfnamefont {V.}~\bibnamefont {Fock}},\
  }\bibfield  {title} {\bibinfo {title} {Beweis des adiabatensatzes},\
  }\href@noop {} {\bibfield  {journal} {\bibinfo  {journal} {Zeitschrift
  f{\"u}r Physik}\ }\textbf {\bibinfo {volume} {51}},\ \bibinfo {pages} {165}
  (\bibinfo {year} {1928})}\BibitemShut {NoStop}%
\bibitem [{\citenamefont {Kato}(1950)}]{kato1950adiabatic}%
  \BibitemOpen
  \bibfield  {author} {\bibinfo {author} {\bibfnamefont {T.}~\bibnamefont
  {Kato}},\ }\bibfield  {title} {\bibinfo {title} {On the adiabatic theorem of
  quantum mechanics},\ }\href@noop {} {\bibfield  {journal} {\bibinfo
  {journal} {Journal of the Physical Society of Japan}\ }\textbf {\bibinfo
  {volume} {5}},\ \bibinfo {pages} {435} (\bibinfo {year} {1950})}\BibitemShut
  {NoStop}%
\bibitem [{\citenamefont {Zhang}\ and\ \citenamefont
  {Gong}(2020)}]{zhang2020dissipative}%
  \BibitemOpen
  \bibfield  {author} {\bibinfo {author} {\bibfnamefont {D.-J.}\ \bibnamefont
  {Zhang}}\ and\ \bibinfo {author} {\bibfnamefont {J.}~\bibnamefont {Gong}},\
  }\bibfield  {title} {\bibinfo {title} {Dissipative adiabatic measurements:
  Beating the quantum cram{\'e}r-rao bound},\ }\href@noop {} {\bibfield
  {journal} {\bibinfo  {journal} {Physical Review Research}\ }\textbf {\bibinfo
  {volume} {2}},\ \bibinfo {pages} {023418} (\bibinfo {year}
  {2020})}\BibitemShut {NoStop}%
\bibitem [{\citenamefont {Piacentini}\ \emph {et~al.}(2017)\citenamefont
  {Piacentini}, \citenamefont {Avella}, \citenamefont {Rebufello},
  \citenamefont {Lussana}, \citenamefont {Villa}, \citenamefont {Tosi},
  \citenamefont {Gramegna}, \citenamefont {Brida}, \citenamefont {Cohen},
  \citenamefont {Vaidman} \emph {et~al.}}]{piacentini2017determining}%
  \BibitemOpen
  \bibfield  {author} {\bibinfo {author} {\bibfnamefont {F.}~\bibnamefont
  {Piacentini}}, \bibinfo {author} {\bibfnamefont {A.}~\bibnamefont {Avella}},
  \bibinfo {author} {\bibfnamefont {E.}~\bibnamefont {Rebufello}}, \bibinfo
  {author} {\bibfnamefont {R.}~\bibnamefont {Lussana}}, \bibinfo {author}
  {\bibfnamefont {F.}~\bibnamefont {Villa}}, \bibinfo {author} {\bibfnamefont
  {A.}~\bibnamefont {Tosi}}, \bibinfo {author} {\bibfnamefont {M.}~\bibnamefont
  {Gramegna}}, \bibinfo {author} {\bibfnamefont {G.}~\bibnamefont {Brida}},
  \bibinfo {author} {\bibfnamefont {E.}~\bibnamefont {Cohen}}, \bibinfo
  {author} {\bibfnamefont {L.}~\bibnamefont {Vaidman}}, \emph {et~al.},\
  }\bibfield  {title} {\bibinfo {title} {Determining the quantum expectation
  value by measuring a single photon},\ }\href@noop {} {\bibfield  {journal}
  {\bibinfo  {journal} {Nature Physics}\ }\textbf {\bibinfo {volume} {13}},\
  \bibinfo {pages} {1191} (\bibinfo {year} {2017})}\BibitemShut {NoStop}%
\bibitem [{\citenamefont {Pan}\ \emph {et~al.}(2020)\citenamefont {Pan},
  \citenamefont {Zhang}, \citenamefont {Cohen}, \citenamefont {Wu},
  \citenamefont {Chen},\ and\ \citenamefont {Davidson}}]{pan2020weak}%
  \BibitemOpen
  \bibfield  {author} {\bibinfo {author} {\bibfnamefont {Y.}~\bibnamefont
  {Pan}}, \bibinfo {author} {\bibfnamefont {J.}~\bibnamefont {Zhang}}, \bibinfo
  {author} {\bibfnamefont {E.}~\bibnamefont {Cohen}}, \bibinfo {author}
  {\bibfnamefont {C.-w.}\ \bibnamefont {Wu}}, \bibinfo {author} {\bibfnamefont
  {P.-X.}\ \bibnamefont {Chen}},\ and\ \bibinfo {author} {\bibfnamefont
  {N.}~\bibnamefont {Davidson}},\ }\bibfield  {title} {\bibinfo {title}
  {Weak-to-strong transition of quantum measurement in a trapped-ion system},\
  }\href@noop {} {\bibfield  {journal} {\bibinfo  {journal} {Nature Physics}\
  }\textbf {\bibinfo {volume} {16}},\ \bibinfo {pages} {1206} (\bibinfo {year}
  {2020})}\BibitemShut {NoStop}%
\bibitem [{\citenamefont {P{\'e}rez-Salinas}\ \emph {et~al.}(2020)\citenamefont
  {P{\'e}rez-Salinas}, \citenamefont {Cervera-Lierta}, \citenamefont
  {Gil-Fuster},\ and\ \citenamefont {Latorre}}]{perez2020data}%
  \BibitemOpen
  \bibfield  {author} {\bibinfo {author} {\bibfnamefont {A.}~\bibnamefont
  {P{\'e}rez-Salinas}}, \bibinfo {author} {\bibfnamefont {A.}~\bibnamefont
  {Cervera-Lierta}}, \bibinfo {author} {\bibfnamefont {E.}~\bibnamefont
  {Gil-Fuster}},\ and\ \bibinfo {author} {\bibfnamefont {J.~I.}\ \bibnamefont
  {Latorre}},\ }\bibfield  {title} {\bibinfo {title} {Data re-uploading for a
  universal quantum classifier},\ }\href@noop {} {\bibfield  {journal}
  {\bibinfo  {journal} {Quantum}\ }\textbf {\bibinfo {volume} {4}},\ \bibinfo
  {pages} {226} (\bibinfo {year} {2020})}\BibitemShut {NoStop}%
\bibitem [{\citenamefont {Schuld}\ \emph {et~al.}(2021)\citenamefont {Schuld},
  \citenamefont {Sweke},\ and\ \citenamefont {Meyer}}]{schuld2021effect}%
  \BibitemOpen
  \bibfield  {author} {\bibinfo {author} {\bibfnamefont {M.}~\bibnamefont
  {Schuld}}, \bibinfo {author} {\bibfnamefont {R.}~\bibnamefont {Sweke}},\ and\
  \bibinfo {author} {\bibfnamefont {J.~J.}\ \bibnamefont {Meyer}},\ }\bibfield
  {title} {\bibinfo {title} {Effect of data encoding on the expressive power of
  variational quantum-machine-learning models},\ }\href@noop {} {\bibfield
  {journal} {\bibinfo  {journal} {Physical Review A}\ }\textbf {\bibinfo
  {volume} {103}},\ \bibinfo {pages} {032430} (\bibinfo {year}
  {2021})}\BibitemShut {NoStop}%
\bibitem [{\citenamefont {Li}\ \emph {et~al.}(2017)\citenamefont {Li},
  \citenamefont {Yang}, \citenamefont {Peng},\ and\ \citenamefont
  {Sun}}]{li2017hybrid}%
  \BibitemOpen
  \bibfield  {author} {\bibinfo {author} {\bibfnamefont {J.}~\bibnamefont
  {Li}}, \bibinfo {author} {\bibfnamefont {X.}~\bibnamefont {Yang}}, \bibinfo
  {author} {\bibfnamefont {X.}~\bibnamefont {Peng}},\ and\ \bibinfo {author}
  {\bibfnamefont {C.-P.}\ \bibnamefont {Sun}},\ }\bibfield  {title} {\bibinfo
  {title} {Hybrid quantum-classical approach to quantum optimal control},\
  }\href@noop {} {\bibfield  {journal} {\bibinfo  {journal} {Physical review
  letters}\ }\textbf {\bibinfo {volume} {118}},\ \bibinfo {pages} {150503}
  (\bibinfo {year} {2017})}\BibitemShut {NoStop}%
\bibitem [{\citenamefont {Alam}\ and\ \citenamefont
  {Ghosh}(2022)}]{alam2022deepqmlp}%
  \BibitemOpen
  \bibfield  {author} {\bibinfo {author} {\bibfnamefont {M.}~\bibnamefont
  {Alam}}\ and\ \bibinfo {author} {\bibfnamefont {S.}~\bibnamefont {Ghosh}},\
  }\bibfield  {title} {\bibinfo {title} {Deepqmlp: A scalable quantum-classical
  hybrid deepneural network architecture for classification},\ }\href@noop {}
  {\bibfield  {journal} {\bibinfo  {journal} {arXiv preprint arXiv:2202.01899}\
  } (\bibinfo {year} {2022})}\BibitemShut {NoStop}%
\bibitem [{\citenamefont {Henderson}\ \emph {et~al.}(2020)\citenamefont
  {Henderson}, \citenamefont {Shakya}, \citenamefont {Pradhan},\ and\
  \citenamefont {Cook}}]{henderson2020quanvolutional}%
  \BibitemOpen
  \bibfield  {author} {\bibinfo {author} {\bibfnamefont {M.}~\bibnamefont
  {Henderson}}, \bibinfo {author} {\bibfnamefont {S.}~\bibnamefont {Shakya}},
  \bibinfo {author} {\bibfnamefont {S.}~\bibnamefont {Pradhan}},\ and\ \bibinfo
  {author} {\bibfnamefont {T.}~\bibnamefont {Cook}},\ }\bibfield  {title}
  {\bibinfo {title} {Quanvolutional neural networks: powering image recognition
  with quantum circuits},\ }\href@noop {} {\bibfield  {journal} {\bibinfo
  {journal} {Quantum Machine Intelligence}\ }\textbf {\bibinfo {volume} {2}},\
  \bibinfo {pages} {1} (\bibinfo {year} {2020})}\BibitemShut {NoStop}%
\bibitem [{\citenamefont {Liu}\ \emph {et~al.}(2021{\natexlab{b}})\citenamefont
  {Liu}, \citenamefont {Lim}, \citenamefont {Wood}, \citenamefont {Huang},
  \citenamefont {Guo},\ and\ \citenamefont {Huang}}]{liu2021hybrid}%
  \BibitemOpen
  \bibfield  {author} {\bibinfo {author} {\bibfnamefont {J.}~\bibnamefont
  {Liu}}, \bibinfo {author} {\bibfnamefont {K.~H.}\ \bibnamefont {Lim}},
  \bibinfo {author} {\bibfnamefont {K.~L.}\ \bibnamefont {Wood}}, \bibinfo
  {author} {\bibfnamefont {W.}~\bibnamefont {Huang}}, \bibinfo {author}
  {\bibfnamefont {C.}~\bibnamefont {Guo}},\ and\ \bibinfo {author}
  {\bibfnamefont {H.-L.}\ \bibnamefont {Huang}},\ }\bibfield  {title} {\bibinfo
  {title} {Hybrid quantum-classical convolutional neural networks},\
  }\href@noop {} {\bibfield  {journal} {\bibinfo  {journal} {Science China
  Physics, Mechanics \& Astronomy}\ }\textbf {\bibinfo {volume} {64}},\
  \bibinfo {pages} {1} (\bibinfo {year} {2021}{\natexlab{b}})}\BibitemShut
  {NoStop}%
\end{thebibliography}
\end{document}